\documentstyle[11pt]{article}
\input amssymb.sty

\title{The Paraconsistent Logic of \\Quantum Superpositions}

\author{{\sc N. da Costa}$^{1}$\ and {\sc C. de Ronde}\ $^{2}$}
\date{}

\begin{document}

\bibliographystyle{plain}
\maketitle

\begin{center}
\begin{small}
1. Universidade Federal de Santa Catarina - Brazil \\
2. Instituto de Filosof\'ia "Dr. A. Korn" \\ 
Universidad de Buenos Aires, CONICET - Argentina \\
Center Leo Apostel and Foundations of  the Exact Sciences\\
Brussels Free University - Belgium \\
\end{small}
\end{center}

\begin{abstract}
\noindent Physical superpositions exist both in classical and in
quantum physics. However, what is exactly meant by `superposition'
in each case is extremely different. In this paper we discuss some
of the multiple interpretations which exist in the literature
regarding superpositions in quantum mechanics. We argue that all
these interpretations have something in common: they all attempt to
avoid `contradiction'. We argue in this paper, in favor of the importance of developing a new interpretation
of superpositions which takes into account contradiction, as a key
element of the formal structure of the theory, ``right from the start''. In order to show the feasibility of our interpretational project we present an outline of a paraconsistent approach to quantum superpositions which attempts to account for the contradictory properties present in general within quantum superpositions. This approach must not be understood as a closed formal and conceptual scheme but rather as a first step towards a different type of understanding regarding quantum superpositions. 
\end{abstract}
\begin{small}



\end{small}

\bibliography{pom}

\newtheorem{theo}{Theorem}[section]

\newtheorem{definition}[theo]{Definition}

\newtheorem{lem}[theo]{Lemma}

\newtheorem{met}[theo]{Method}

\newtheorem{prop}[theo]{Proposition}

\newtheorem{coro}[theo]{Corollary}

\newtheorem{exam}[theo]{Example}

\newtheorem{rema}[theo]{Remark}{\hspace*{4mm}}

\newtheorem{example}[theo]{Example}

\newcommand{\proof}{\noindent {\em Proof:\/}{\hspace*{4mm}}}

\newcommand{\qed}{\hfill$\Box$}

\newcommand{\ninv}{\mathord{\sim}} 

\newtheorem{postulate}[theo]{Postulate}

\section{Introduction}

There is an important link in the history of physics between the interpretation of theories and their formal development. Relativity would have not been possible without non-euclidean geometry nor classical physics without infinitesimal calculus. In quantum mechanics, the formal scheme was elaborated ---mainly by Schr\"odingier, Heisenberg, Born, Jordan and Dirac--- almost in parallel to the orthodox interpretation. However, still today, the interpretation of quantum mechanics remains controversial regarding most of its non-classical characteristics: indeterminism, holism, contextuality, non-locality, etc. There is, still today, no consensus regarding the meaning of such expressions of the theory. In this paper we shall be concerned with a specific aspect of quantum mechanics, namely, the principle of superposition which, as it is well known, gives rise to the so called quantum superpositions.  Physical superpositions exist both in classical and in quantum
physics. However, what is exactly meant by ``superposition" in each
case is extremely different. In classical physics one can have
superpositions of waves or fields. A wave (field) $\alpha$ can be
added to a different wave (field) $\beta$ and the sum will give a
`new' wave (field) $\mu = \alpha + \beta$. There is in this case no
weirdness for the sum of multiple states gives as a result a new
single state. In quantum mechanics on the contrary, a linear combination of multiple
states, $\alpha + \beta$ is not reducible to one single state, and there is no
obvious interpretation of such superposition of states. As a matter of fact, today
quantum superpositions play a central role within the most outstanding
technical developments such as quantum teleportation, quantum
cryptography and quantum computation \cite{Nature05, Nature07}. The
question we attempt to address in this paper regards the meaning  and physical representation of quantum superpositions. There are
many interpretations of quantum mechanics each of which provides an
answer to this question. In the following we shall review some of
these proposals. We shall then argue in favor of the possibility to
develop a new interpretation which considers contradictory properties as a main aspect of quantum superpositions and present an outline of a formal approach based on paraconsistent logic which attempts to consider contradiction  ``right from the start''. We must remark that we do not understand this approach as a closed formal and conceptual scheme but rather as a first step towards a different type of understanding regarding quantum superpositions. 

Paraconsistent logics are the logics of inconsistent but nontrivial
theories. The origins of paraconsistent logics go back to the first
systematic studies dealing with the possibility of rejecting the
principle of noncontradiction. Paraconsistent logic was elaborated,
independently, by Stanislaw Jaskowski in Poland, and by the first
author of this paper in Brazil, around the middle of the last
century (on paraconsistent logic, see, for example:
\cite{daCostaKrauseBueno07}). A theory $T$ founded on the logic $L$,
which contains a symbol for negation, is called inconsistent if it
has among its theorems a sentence $A$ and its negation $\neg A$;
otherwise, it is said to be consistent. $T$ is called trivial if any
sentence of its language is also a theorem of $T$; otherwise, $T$ is
said to be non-trivial. In classical logics and in most usual
logics, a theory is inconsistent if, and only if, it is trivial. $L$
is paraconsistent when it can be the underlying logic of
inconsistent but non trivial theories. Clearly classical logic and all usual logics are not paraconsistent. 
The importance of paraconsistent logic is not
limited to the realm of pure logic but has been extended to many
fields of application such as robot control, air traffic control
\cite{NakamatsuAbeSuzuki02a}, control systems for autonomous
machines \cite{NakamatsuAbeSuzuki02b}, defeasible deontic reasoning \cite{NakamatsuAbeSuzuki01},
information systems \cite{Akama01} and medicine. 

In the following, we attempt to call the attention to the importance of extending the
realm of paraconsistent logic to the formal account of quantum
superpositions. Firstly, we shall discuss the very different meanings of the
term `superposition' in both classical and quantum physics. In
section 3, we shall present some of the very different interpretations of the meaning of a quantum
superposition which can be found in the literature. In section 4, we shall argue in favor of
the importance of considering an interpretation of superposition in
terms of paraconsistent logic. In section 5, we present a formal scheme in terms of paraconsistent logic which attempts to account for the inner contradictions present within a quantum superposition. Finally, in section 6, we argue in favor of considering contradiction ``right from the start''.

\section{What is a Quantum Superposition?}

In classical physics, every physical system may be described
exclusively by means of its \emph{actual properties}, taking
`actuality' as expressing the \emph{preexistent} mode of being of
the properties themselves, independently of observation ---the `pre'
referring to its existence previous to measurement. Each system has
a determined state characterized mathematically in terms of a point
in phase space. The change of the system may be described by the
change of its actual properties. Potential or possible properties
are considered as the points to which the system might arrive in a
future instant of time. The occurrence of possibilities in such cases merely reflects our
ignorance about what is actual. Contrary to what seems to happen in quantum mechanics, statistical states do not
correspond to features of the actual system, but quantify our lack of
knowledge of those actual features (\cite{Dieks10}, p.
125). 

Classical mechanics tells us via the equation of motion how the
state of the system moves along the curve determined by initial
conditions in the phase space. The representation of the state of
the physical system is given by a point in phase space $\Gamma$ and
the physical magnitudes are represented by real functions over
$\Gamma$. These functions commute between each other and can be
interpreted as possessing definite values independently of physical
observation, i.e. each magnitude can be interpreted as being
actually preexistent to any possible measurement. In the orthodox
formulation of quantum mechanics, the representation of the state of
a system is given by a ray in Hilbert space ${\cal H}$. But,
contrary to the classical scheme, physical magnitudes are
represented by operators on ${\cal H}$ that, in general, do not
commute. This mathematical fact has extremely problematic
interpretational consequences for it is then difficult to affirm
that these quantum magnitudes are \emph{simultaneously preexistent}.
In order to restrict the discourse to different sets of commuting
magnitudes, different Complete Sets of Commuting Operators (CSCO)
have to be chosen. The choice of a particular representation (given
by a CSCO) determines the basis in which the observables diagonalize
and in which the ray can be expressed. Thus, the ray can be written
as different linear combinations of states:

\begin{equation}\label{cl}
\alpha_{i}|\varphi_{i}^{B1}> + \alpha_{j}|\varphi^{B1}_j> =
|\varphi^{B2}_q> =\beta_{m}|\varphi_{m}^{B3}> + \beta_{n}
|\varphi_{n}^{B3}> + \beta_{o} |\varphi_{o}^{B3}>
\end{equation}

\noindent The linear combinations of states are also called quantum
superpositions. As it was clearly expressed by Dirac \cite{Dirac74}: 
``The nature of the relationships which the superposition
principle requires to exist between the states of any system is of a
kind that cannot be explained in terms of familiar physical
concepts. One cannot in the classical sense picture a system being
partly in each of two states and see the equivalence of this to the
system being completely in some other state." The formal difference of using vectors in ${\cal H}$ instead of
points in $\Gamma$ seems to imply that in quantum mechanics
---apart from the `possibility' which is encountered in classical
mechanics--- there is another, different realm which must be
necessarily considered and refers, at each instant of time, to {\it
contradictory properties}. To see this, consider the following
example: given a spin $1/2$ system whose state is $|\uparrow_{z}>$,
we let it interact with a magnetic field in the $z$ direction. All
outcomes that can become actual in the future are potential
properties of the system, in an analogous manner as all possible
reachable positions of a pendulum are in the classical case. But at
each instant of time, for example at the initial instant, if we
consider the $z$ direction and the projection operator
$|\uparrow_{z}><\uparrow_{z}|$ as representing a preexistent actual
property, there are other incompatible properties arising from
considering projection operators of spin projections in other
directions. For example, in the $x$ direction, the projection
operators $|\uparrow_{x}><\uparrow_{x}|$ and
$|\downarrow_{x}><\downarrow_{x}|$ do not commute with
$|\uparrow_{z}><\uparrow_{z}|$ and thus, cannot be considered to
possess definite values simultaneously. Since Born interpretation of
the wave function, these properties are usually considered as {\it
possible}. However, this possibility is essentially different from
the idea of possibility discussed in classical physics which relates
to the idea of a {\it process}. If we consider that the formalism of
quantum mechanics provides a description of the world, a
representation of {\it what there is}
---and does not merely make reference to measurement outcomes---, at
each instant of time the properties, $|\uparrow_{z}><\uparrow_{z}|$,
$|\uparrow_{x}><\uparrow_{x}|$ and
$|\downarrow_{x}><\downarrow_{x}|$ must be taken into account independently of their future actualization for
they all provide non trivial information about the state of affairs.
In particular, the properties $|\uparrow_{x}><\uparrow_{x}|$ and
$|\downarrow_{x}><\downarrow_{x}|$, which constitute the
superposition and must be considered simultaneously are in general {\it
contradictory properties}.

In the quantum logic approach one of the properties, namely, the one
in which we can write the state of affairs as a single term, is
considered as `actual' while the others are taken to be `potential'
properties. Potential properties can become actual. These properties, e.g.
$|\uparrow_{x}><\uparrow_{x}|$, $|\downarrow_{x}><\downarrow_{x}|$,
$|\uparrow_{y}><\uparrow_{y}|$ and
$|\downarrow_{y}><\downarrow_{y}|$ in our example, are always part
of superpositions with more than one term and are constituted by
contradictory properties. However, from a mathematical perspective,
independently of their mode of existence, both potential and actual
properties are placed at the same level in the algebraic frame which
describes the state of affairs according to quantum mechanics: the
projections of the spin in all directions are atoms of the lattice
and there is no formal priority of the actual over the potential
properties. This rises the question if one can consider quantum superpositions as preexistent entities, independently of their future actualization.  

In the laboratory, it is precisely this contradictory
potential realm which is necessary to be considered by the
experimentalist in the developments which are taking place today
regarding the processing of quantum information as quantum computing
and quantum communication \cite{Nature05, Nature07}. This seems to
point in the direction that these properties have an existence which
cannot be reduced to their {\it becoming actual} at a future instant
of time. Superpositions correspond to possible outcomes which occur on 
an equal footing in the superposition of the final state, so that there is no sign that any one of them is
more real than any other (\cite{Dieks10}, p. 120).
Taking these problems into account there are many interpretations which attempt
to provide an answer to the question: what is a quantum superposition? We shall discuss in the
next section some of these proposals.

\section{The Multiple Interpretations of Quantum Superpositions}

As we have seen above, the formal description of quantum mechanics
seems to imply a deep departure from the classical notion of
possible or probable. This was cleverly exemplified by Erwin Schr\"odinger in his famous cat experiment  \cite{Schrodinger35}, in which a half dead and half alive cat seemed to laugh of the idea of possessing a determined state. However, one can find in the literature, there are many different
interpretations of quantum mechanics in general and of the meaning
of a quantum superposition in particular. In this section
we shall review some of these very distinct interpretations. We do not attempt to provide a complete review of interpretations but rather to analyze instead their specific understanding of quantum superpositions.  

Although we must take into account the fact that 'state vector' and 'cat' are two concepts in different levels of discourse (\cite{Dieks88a}, p. 189). From a realist perspective, which considers physics as providing a description or an
expression of the world, the question still arises, if this formal
or mathematical representation given by superpositions, namely
equation \ref{cl}
---which allow us to calculate the probability of the possible
measurement outcomes---, can be related conceptually to a notion
which can allow us to think, independently of measurement outcomes,
about the `superposition of states in Hilbert space' in an analogous
manner as we think of a `point in phase space' (in the formal level)
as describing an `object in space-time' (in the conceptual level).
What is describing a mathematical superposition? Can we create or
find adequate concepts which can provide a representational
realistic account of a quantum superposition independent of
measurement outcomes? Of course, from a general
empiricist perspective one is not committed to answering these set
of questions. The idea that the quantum wave function as related to a
superposition is just a theoretical device with no ontological
content goes back to Bohr's interpretation of quantum mechanics. The
impossibility to interpret the quantum wave function in an
ontological fashion can be understood in relation to his
characterization of $\Psi$ in terms of an algorithmic device which
computes measurement results.\footnote{According to Bohr
(\cite{VW74}, p. 338) the Schr\"odinger wave equation is just an
abstract method of calculus and it does not designate in itself
any phenomena. See also \cite{BokulichBokulich} for discussion.} This position radically addressed seems to end up in
the instrumentalistic account shared implicitly by many and
developed explicitly by Fuchs and Peres \cite{FuchsPeres00}. Bas van
Fraassen, whom we consider a close follower of Bohr's ideas, has
also taken an anti-metaphysical position with respect to the
interpretation of the quantum wave function. His justification
stands on his empiricist account of both physics and philosophy (see
\cite{VF91}, section 9.1). 

From an empiricist perspective the formalism does not provide a
description of {\it what there is}. Superpositions are thus, a
theoretical device through which one can consider the actual
observation {\it hic et nunc}. Empiricism can be linked to
probability in terms of the frequency interpretation which rests,
contrary to the original conception of probability, not on the idea
that probability describes in terms of ignorance an existent state
of affairs, but rather in a set of empirical results found in a
series of measurements. However, and independently of the problems
encountered within such empiricist stances, if superpositions are
considered just as a {\it theoretical device}, then the question of
interpretation seems to loose its strength. For why should we pursue
an interpretation if, like Fuchs and Peres remark, quantum mechanics
does the job and already provides an algorithm for computing
probabilities for the macroscopic events? There are other reasons
which one could put forward to account for the importance of
interpretation even from an empiricist perspective (see for example
van Fraassen \cite{VF80}), however these reasons must remain only secondary
in the quest of science.

On the contrary, from a realist position, there is need to provide an answer to the link between the theory and its conceptual understanding of the world. To put it in  a nutshell: what is quantum mechanics telling us about the world?
As noticed by Bacciagaluppi (\cite{Bacciagaluppi96}, p. 74), the
hidden variable program attempts to ``restore a classical way of
thinking about {\it what there is}.'' In this sense, Bohm's proposal
seems to restore the possibility of discussing in terms of a state of affairs described in terms of a
set of definite valued properties. In Bohmian mechanics the state of
a system is given by the wave function $\Psi$ together with the
configuration of particles $X$. The quantum wave function must be
understood in analogy to a classical field that moves the particles
in accordance with the following functional relation: $\frac{dx}{dt}
= \nabla S$, where $S = \hbar \delta$ ($\delta$ being the phase of
$\psi$). Thus, particles always have a well defined position
together with the rest of their properties and the evolution depends
on the quantum field. It then follows that, there are no
superpositions of states, the superposition is given only at the
level of the field and remains as mysterious as the superposition of
classical fields. The field does not
only have a dynamical character but also determines the epistemic
probability of the configuration of particles {\it via} the usual
Born rule.

A different approach, which starts from a particular interpretation of quantum superpositions is the so called many worlds interpretation (MW),  
considered to be a direct conclusion from
Everett's first proposal in terms of `relative states'
\cite{Everett57}. Everett's idea was to let quantum mechanics find
its own interpretation, making justice to the symmetries inherent in
the Hilbert space formalism in a simple and convincing way
\cite{DeWittGraham}. MW interpretations are no-collapse interpretations which
respect the orthodox formulation of quantum mechanics. The main idea behind many worlds
interpretations is that superpositions relate to collections of
worlds, in each of which exactly one value of an observable, which
corresponds to one of the terms in the superposition, is realized.
Apart from being simple, the claim is that it possesses a natural
fit to the formalism, respecting its symmetries. The solution
proposed to the measurement problem is provided by assuming that
each one of the terms in the superposition is {\it actual} in its
own correspondent world. Thus, it is not only the single value which we see in `our world'
which gets actualized but rather, that a branching of worlds takes
place in every measurement, giving rise to a multiplicity of worlds
with their corresponding actual values. The possible splits of the
worlds are determined by the laws of quantum mechanics but each
world becomes again `classical'. Quantum superpositions are interpreted as expressing the existence
of multiple worlds, each of which exists in (its own) actuality.
However, there are no superpositions in this, our actual world, for
each world becomes again a ``classical world''. The many worlds
interpretation seems to be able to recover these islands of
classicality at the price of multiplying the `actual realm'. In this
case, the quantum superposition is expelled from each actual world
and recovered only in terms of the relation between the multiple
worlds.

The Geneva school to quantum logic and similar approaches such 
as that of Foulis and Randall \cite{FoulisPironRandall83} attempt to
consider quantum physics as related to the realms of actuality and
potentiality in analogous manner to classical physics. According to
the Geneva school, both in classical and quantum physics
measurements will provoke fundamental changes of the state of the
system.\footnote{What is special for a classical system, is that
`observables' can be described by functions on the state space. This
is the main reason that, a measurement corresponding to such an
observable, can be left out of the description of the theory `in
case one is not interested in the change of state provoked by the
measurement', but `only interested in the values of the
observables'. It is in this respect that the situation is very
different for a quantum system. Observables can also be described,
as projection valued measures on the Hilbert space, but `no definite
values can be attributed to such a specific observable for a
substantial part of the states of the system'. For a quantum system,
contrary to a classical system, it is not true that `either a
property or its negation is actual'.} Continuing Heisenberg's considerations
in the new physics, Constantin Piron has been one of the leading
figures in developing the notion of potentiality within the logical
structure of quantum mechanics \cite{Piron76, Piron83}. Following
\cite{Smets05}, a physical property, never mind whether a classical
or quantum one, is specified as what corresponds to a set of
definite experimental projects. A {\it definite experimental
project} (DEP) is an experimental procedure (in fact, an equivalence
class of experimental procedures) consisting in a list of actions
and a rule that specifies in advance what has to be considered as a
{\it positive} result, in correspondence with the {\it yes} answer
to a dichotomic question. Each DEP tests a property. A given DEP is called {\it certain}
(correspondingly, a dichotomic question is called {\it true}) if it
is sure that the positive response would be obtained when the
experiment is performed or, more precisely, in case that whenever
the system is placed in a measurement situation then it produces
certain definite phenomenon to happen. A physical property is called
{\it actual} in case the DEPs which test it are certain and it is
called {\it potential} otherwise. Whether a property is actual or
potential depends on the state in which one considers the system to
be. Though in this approach both actuality and potentiality are
considered as modes of being, actual properties are considered as
attributes that {\it exist}, in the EPR sense as  elements of physical reality\footnote{Einstein designed, in the by now famous EPR `paper' \cite{EPR}, a definition of when a physical quantity could be considered an  {\it element of physical reality} within quantum mechanics. By using this definition Einstein, Podolsky and Rosen argued against the completeness of the quantum theory. For a general discussion see \cite{EPRStanford}.},
while potential properties are not conceived as existing in the same
way as real ones. They are thought as {\it possibilities} with
respect to actualization, because potential properties may be
actualized due to some change in the state of the system. In this
case the superposition provides a measure ---given by the real
numbers which appear in the same term as the state--- over the
potential properties which could become actual in a given
situation.

\section{Quantum Superpositions and the `Contradiction' of Properties?}

Although the interpretations we have discussed in the previous
section from both their formal and metaphysical commitments have
many differences, there is still something they all share in common:
they all attempt to avoid contradictions. Indeed `contradiction' has
been regarded with disbelief in Western thought due to certain
metaphysical presuppositions which go back to Plato, Aristotle,
Leibniz and Kant. Even after the development of paraconsistent logic
in the mid XX century and the subsequent technical progress
this theory has allowed, the aversion towards contradiction is
still present today within science and philosophy. The famous statement of Popper that the
acceptance of inconsistency ``would mean the complete breakdown of
science" remains an unfortunate prejudice within present philosophy
of science (see \cite{daCostaFrench03}, Chap. 5).

Leaving instrumentalist positions aside, one of us has argued
elsewhere \cite{deRonde10} that one can find in the vast literature
regarding the interpretation of quantum mechanics, two main
strategies which attempt to provide an answer to the riddle of `what is
quantum mechanics talking about'. The first strategy is to begin
with a presupposed set of metaphysical principles and advance
towards a new formalism. Examples of this strategy are Bohmian
mechanics, which has been discussed above, and the collapse theory proposed by Ghirardi, Rimini and Weber (also called `GRW theory') \cite{GRW}, which
introduces non-linear terms in the Schr\"odinger equation. The
second strategy is to accept the orthodox formalism of quantum
mechanics and advance towards the creation and elucidation of the
metaphysical principles which would allow us to answer the question:
`what is quantum mechanics talking about'? Examples of this second
strategy are quantum logic and its different lines of development
such as the just described Geneva School of Jauch and Piron, and the modal
interpretation (see for example \cite{DicksonDieks02, deRonde11, Vermaas99}). From this perspective, the importance is to focus in
the formalism of the theory and try to learn about the symmetries,
the logical features and structural relations. The idea is that, by
learning about such aspects of the theory we can also develop the
metaphysical conditions which should be taken into account in a
coherent ontological interpretation of quantum mechanics.

But even independently of the choice of this strategy, it seems quite clear that technical developments which are taking place today regarding quantum mechanics have advanced quite independently of the commitments to any classical metaphysical background. Quantum computation makes use of the multiple flow of information in
the superposition even considering (in principle) contradictory
paths. Also quantum cryptography uses the relation between
contradictory terms in order to send messages avoiding classical
spies. At a formal level, the path integral approach takes into account
the multiple contradictory paths within two points
\cite{FeynmanHibbs65}. Thus, since both the formalism and experiments seem
to consider `contradictory elements' within quantum mechanics, we
argue that it can be of deep interest to advance towards a formalism which
takes contradiction into account ``right from the
start".\footnote{In an analogous fashion as D\'ecio Krause has
developed a Q-set theory which accounts for indistinguishable
particles with a formal calculus ``right from the start"
\cite{Krause92}.} Evidently, such a formalism could open paths not only to continue the technical developments just mentioned but also to understand the meaning of quantum superpositions from a new perspective. Our proposal is twofold, firstly, to call the attention of the importance of considering contradictory properties within the formalism and interpretation of quantum superpositions; and secondly, to show that paraconsistent logics can open a formal line of research.  In the next section we make a first step in this same direction, providing an outline to an approach based on paraconsistent logic.

\section{An Outline of a Paraconsistent Approach to Quantum Superpositions}

We bring into, now, a paraconsistent logical system $ZF_{1}$, that
is a strong set theory, even stronger than common ZF (Zermelo
Frenkel set theory). On $ZF_{1}$ and related matters, see
\cite{daCostaKrauseBueno07}. In what follows, we employ the
terminology, notations and conventions of Kleene \cite{Kleene64} 

The basic symbols of the langugae $ZF_{1}$ are the following: 1)
Propositional connectives: implication ($\rightarrow$), conjunction
($\wedge$), disjunction ($\vee$) and (weak) negation ($\neg$),
equivalence ($\leftrightarrow$) is defined as usual. 2) Individual
variables: a demmuerable set of variables, that are represented by
smal Latin letters of the end of the aplphabet. 3) The quantifiers
$\forall$ (for all) and $\exists$ (there exists). 4) The binary
predicate symbols $\in$ (membership) and $=$ (identity). 5)
Auxiliary symbols: parenthesis.

Syntactic notions, for example those of formula, closed formula or
sentence, and free occurrence of a variable in a formula, are
defined as customary. Russell's symbol for description ($\iota$) is
introduced by contextual definition and with the help of the
description, the classifier $\{ x; F(x) \}$, where $F(x)$ is a
formula and $x$ a variable.

\begin{definition}
{\rm $A^{\circ}$ abbreviates $\neg(A \wedge \neg A)$.}
\end{definition}

Loosely speaking, $A^{\circ}$ means that $A$ is a well-behaved
formula, i.e., that it is not the case that one has $A$ and $\neg A$
both true (or, what is the same thing, that the contradiction $A
\wedge \neg A$ is false).

\begin{definition}
{\rm $\neg^{*}A$ abbreviates $\neg A \wedge A^{\circ}$.}
\end{definition}

$\neg^{*}$ functions like a strong negation, a kind of classical
negation in our logic. On the other hand, $\neg$ is the weak (or
paraconsistent) negation.\\

\textbf{Postulates of $ZF_{1}$}\\

\emph{a) Propositional postulates:}

\begin{enumerate}
\item
$A \rightarrow (B \rightarrow A)$

\item
$(A \rightarrow B) \rightarrow (A \rightarrow (B \rightarrow C))
\rightarrow (A\rightarrow C)$

\item
$\frac{A {\rm  \ } {\rm  \ } {\rm  \ }  {\rm  \ } A \rightarrow B}{B}$

\item
$(A \wedge B) \rightarrow A$

\item
$(A \wedge B) \rightarrow B$

\item
$A \rightarrow (B \rightarrow (A \wedge B))$

\item
$A \rightarrow (A \vee B)$

\item
$B \rightarrow (A \vee B)$

\item
$(A \rightarrow C) \rightarrow ((B\rightarrow C) \rightarrow ((A
\vee B) \rightarrow C))$

\item
$A \vee \neg A$

\item
$\neg \neg A \rightarrow A$

\item
$B^{\circ} \rightarrow ((A \rightarrow B) \rightarrow ((A
\rightarrow \neg B) \rightarrow \neg A))$

\item
$(A^{\circ} \wedge B^{\circ}) \rightarrow (A \rightarrow B)^{\circ}$

\item
$(A^{\circ} \wedge B^{\circ}) \rightarrow (A \wedge B)^{\circ}$

\item
$(A^{\circ} \wedge B^{\circ}) \rightarrow (A \vee B)^{\circ}$
\end{enumerate}

\emph{b) Quantificational Postulaes:}

\begin{enumerate}
\item
$\frac{C \rightarrow A(x)}{C \rightarrow \forall x A(x)}$

\item
$\forall x A(x) \rightarrow A(t)$

\item
$A(t) \rightarrow \exists x A(x)$

\item
$\frac{A(x) \rightarrow C}{\exists x A(x) \rightarrow C}$

\item
$\forall x (A(x))^{\circ} \rightarrow (\forall x A(x))^{\circ}$

\item
$\forall x (A(x))^{\circ} \rightarrow (\exists x A(x))^{\circ}$

\noindent The preceding postulates are subject to the usual
restrictions. In classical logic, as well as in most usual logics, 
we are allowed to reletter bind variables and suppress vacuous variables, 
but it seems that this is not probable in connection with our system. Therefore, we introduce the postulate:

\item
If $B$ is a formula obtained from $A$ by relettering bound variables
or by the suppression of void quantifiers, then $A \leftrightarrow
B$ is an axiom.
\end{enumerate}

\noindent \textbf{Remark.} Postulates 1-15 constitute a propositional system of paraconsistent logic  and adding
the quantificational postulates we obtain a first-order quantificational paraconsistent logic.\\

\emph{c) Set-Theoretic Postulates:}

They are all those of classical $ZF$ in whose formulations the symbol
of negation is replaced by the symbol of strong negation $\neg*$.
The postulates can be formulated supposing that $ZF_{1}$ is a pure
set theory or a theory that contains \textit{Urelemente} (objects that are
not sets). Our results don't depend on the version employed of
$ZF_{1}$. Moreover the existence of some sets that cause problems
(do not exist) in $ZF$, like Russell's collection, could be
postulated as existing in $ZF_{1}$; however, this possibility is
here excluded. ($ZF$ is studied, for instance, in
\cite{Mendelson10}).\\

From now on, capital letters stand for formulas. We have (see
\cite{daCostaKrauseBueno07}):

\begin{definition}
{\rm $\vdash A$ stands for $A$ is a theorem of $ZF_{1}$; $\nvdash A$
is the negation of $\vdash A$.}
\end{definition}

\begin{definition}
{\rm $A^{*}$ is the formula obtained from A by replacing any
occurrence of $\neg$ by an occurrence of $\neg^{*}$}.
\end{definition}

\begin{theo}
In $ZF_{1}$:\\ $\vdash A \vee \neg A, \vdash \neg \neg A \rightarrow A, \vdash (A \wedge \neg^{*} A) \rightarrow B,
\nvdash (A \wedge \neg A) \rightarrow B, \nvdash A \rightarrow \neg \neg A, \nvdash \neg (A \wedge  \neg A)$.
\end{theo}

\begin{theo}
If $A$ is provable in $ZF$, then $A^{*}$ is probable in $ZF_{1}$ ($ZF$ is
included in $ZF_{1}$).
\end{theo}

\begin{theo}
$ZF$ is inconsistent if and only if $ZF_{1}$ is non trivial.
\end{theo}

$ZF$ is, in a certain sense, contained in $ZF_{1}$. So, in $ZF_{1}$
it is possible to systematize extant classical mathematics; in
consequence, $ZF_{1}$ encompasses all mathematical analysis required
for the treatment of standard quantum mechanics, and this treatment
is similar to the one with classical logic (and set theory) as the
basic logic.

In the study of a quantum system $S$ in $NF_{1}$, we note that its
states behave as in classical quantum mechanics, in the sense that

\begin{equation}
ZF_{1} \nvdash \exists s (s \in \hat{S} \wedge \neg^{*} (s \in
\hat{S}))
\end{equation}

\noindent and some other similar formulas are (apparently) not
probable, where $\hat{S}$ is the set of the states of $S$ included in a given superposition.

When $S$ is in the state of superposition of, say, the states
$s_{1}$ and $s_{2}$ (classically inconsistent), we introduce in
$ZF_{1}$ the extra predicate K and expand the system with the
postulates

\begin{equation}
K(S, s_{1})\ {\rm and}\ \neg K(S, s_{1})
\end{equation}

\noindent as well as:

\begin{equation}
K(S, s_{2})\ {\rm and}\ \neg K(S, s_{2})
\end{equation}

Informally, for instance $K(S_{1}, s_{1})$ means that ``$S$ has the
superposition predicate associated to $s_{1}$" (or the
``paraconsistent predicate associated to $s_{1}$"). In other words,
superposition creates a contradictory situation, giving rise to
contradictory relations. In $ZF_{1}$, we can not directly assume that the linear combination of two classically
incompatible states is an `inconsistent'  state; this is so  because the mathematics of usual quantum
mechanics is classical, and such kind of inconsistency would make our system trivial.

To cope with this situation, we appeal to a new postulate:\\

\noindent\textbf{Postulate of inconsistency}. Let $S$ be a quantum
system which is in the superposition of the (classically
incompatible) states $s_{1}$ and $s_{2}$. Under the hipothesys, we
have:

\begin{equation}
K(S, s_{1}) \wedge \neg K(S, s_{1}) \wedge K(S, s_{2}) \wedge \neg
K(S, s_{2})
\end{equation}

\noindent This means that superposition implies contradiction.
Similarly when superposition involves more than two states.\\

Therefore, $ZF_{1}$ constitutes the underlaying logic of an
inconsistent, but apparently non trivial, quantum mechanics that we
denote by $QM_{1}$; usual quantum mechanics will be denoted by $QM$.
Thus, $QM$ is in a certain sense, contained in $QM_{1}$. But the details of the construction of a paraconsistent quantum
mechanics will be left to a future series of technical works.

\section{Discussion: Considering Contradiction ``Right from the Start''}

Our proposal focuses on the idea that it would be worthwhile to
develop a new interpretation of quantum superpositions which considers contradiction ``right from the start''. We have provided an outline of a paraconsistent approach to quantum superpositions which shows the possibility to consider contradictions also from a formal perspective. However, it should be clear that we do not
take paraconsistent logic to be the ``true logic" which should
replace classical logic; in the same way as we do not regard quantum
mechanics as a theory that should replace classical mechanics
\cite{daCostaFrench03, deRonde11}. From our perspective we argue that physicists should
recognize the possibility to use new forms of logic
---such as paraconsistent logic--- which might help us understanding
features of different domains of reality; features which might not be necessarily
accommodated by means of classical logic. We do not believe there is
a ``true logic", but rather that distinct logical systems can be of
use to develop and understand complementary aspects of reality.
Recalling the words of Albert Einstein: ``It is only the theory
which can tell you what can be observed"\footnote{These words,
according to Heisneberg himself, led him to the development of the
indetermination principle in his foundational paper of 1927.} it
could be argued that only within a theory it is possible to consider
and account for phenomena. From this standpoint the development of
the formalism can be regarded not only as a merely technical
improvement, but also as a way to open new paths of understanding
and even of development of new phenomena. Formal development is not
understood here as going beyond the theory, as improving and showing
something that ``was not there before" in the formalism ---as it is
the case of the GRW theory or Bohmian mechanics. Rather, this development is
understood as taking seriously the features which the theory seems
to show us, exposing them in all their strength, ``right from the
start".

We also have to stress that non relativistic quantum mechanics, based on classical logic and
on the common specific postulates, seems to be consistent in the strict logical meaning. So, a paraconsistent version of it has to postulate, in some way or other, the inconsistent character of determinable situations. It appears to be that there is no possibility that someone can ``deduce", employing any one of a majority of extant logics, contradictory consequences of the specific axioms of non relativistic quantum mechanics.

Evidently, we have to develop and to explore the ideas here
sketched. One of the important points to take into consideration is
that there are numerous ways to obtain such kind of inconsistent
quantum mechanics. In addition, we should verify if there are really
important new results of $QM_{1}$ which are not valid in $QM$. The
philosophical meaning of $QM_{1}$ also deserves detailed analysis.

\section*{Acknowledgments}

The authors wish to thank an anonymous referee for his/her careful reading of our manuscript and useful comments.

\end{document}